\documentclass[aps,pre,twocolumn,superscriptaddress,showkeys,showpacs]{revtex4}

\usepackage{amsmath}
\usepackage{amssymb}
\usepackage[dvips]{graphicx}
\usepackage{colordvi}
\usepackage{times}

\newcommand{\ave}[1]{\langle #1\rangle}

\newcommand{\bea}{\begin{eqnarray}}
\newcommand{\eea}{\end{eqnarray}}
\newcommand{\be}{\begin{equation}}
\newcommand{\ee}{\end{equation}}
\newcommand{\beas}[1]{\begin{subequations}\label{#1}\bea}
\newcommand{\eeas}{\eea\end{subequations}}

\newcommand{\Lpp}{L_\textrm{pp}}
\newcommand{\Ree}{R_\textrm{ee}}
\newcommand{\Ne}{N_e}
\newcommand{\Nee}{{\cal N}_e}
\newcommand{\ZZ}{\ave{Z}}
\newcommand{\row}[7]{$#1$ & $#2$ & $#6$ & $#5$ & $#4$ & $#3$ & $#7$ }
\newcommand{\BBlue}[1]{{\bf #1}}

\begin{document}


\newcommand{\TABLECHAINDIMS}{
\begin{table*}[bthp]
 \begin{ruledtabular}
 \begin{tabular}{rcccccc}
 $N$ & $\ave{\Ree^2}$ & $\ave{\Lpp}^2_\textrm{PPA}$& $\ave{\Lpp^2}_\textrm{PPA}$ & $\ave{\Lpp}^2_\textrm{Z1}$& $\ave{\Lpp^2}_\textrm{Z1}$  & $\ave{Z}_\textrm{Z1}$   \\
 \hline
 \row{20 }{ 29.24 }{ 32.16 }{ 28.21 }{ 37.56 }{ 33.18 }{ 0.127}\\
 \row{28 }{ 42.85 }{ 50.32 }{ 44.03 }{ 58.66 }{ 51.86 }{ 0.287}\\
 \row{35 }{ 54.69 }{ 69.00 }{ 60.52 }{ 79.85 }{ 71.01 }{ 0.462}\\
 \row{50 }{ 80.30 }{ 113.9 }{ 100.2 }{ 129.8 }{ 116.2 }{ 0.823}\\
 \row{70 }{ 114.9 }{ 190.7 }{ 169.7 }{ 213.4 }{ 193.3 }{ 1.337}\\
 \row{100 }{ 169.1 }{ 334.8 }{ 301.8 }{ 373.3 }{ 343.0 }{ 1.995}\\
 \row{125 }{ 215.2 }{ 475.4 }{ 431.9 }{ 522.1 }{ 483.7 }{ 2.514} \\
 \row{140 }{ 233.0 }{ 576.6 }{ 528.1 }{ 633.2 }{ 593.5 }{ 2.876}\\
 \row{175 }{ 289.5 }{ 831.1 }{ 766.5 }{ 900.2 }{ 847.9 }{ 3.541}\\
 \row{250 }{ 421.9 }{ 1577 }{ 1481 }{ 1716 }{ 1646 }{ 5.089}\\
 \row{350 }{ 609.4 }{ 2907 }{ 2764 }{ 3245 }{ 3143 }{ 7.168}\\
 \row{500 }{ 831.0 }{ 5738 }{ 5527 }{ 6188 }{ 6050  }{ 10.261}\\
 \row{700 }{ 1203 }{ 1.084\times10^4 }{ 1.057\times10^{4} }{1.189\times10^{4} }{ 1.170\times10^{4}  }{ 14.343}\\
 \row{875 }{ 1521 }{ 1.659\times10^{4} }{ 1.624\times10^{4} }{ 1.779\times10^{4} }{ 1.757\times10^{4} }{ 17.793}\\
 \row{1750 }{ 3003 }{ 6.294\times10^{4} }{ 6.215\times10^{4} }{ 6.806\times10^{4} }{   6.769\times10^{4} }{ 35.204}\\
 \row{3500 }{ 6157 }{ 2.457\times10^{5} }{ 2.441\times10^{5} }{ 2.599\times10^{5} }{ 2.591\times10^{5} }{ 70.444}
 \end{tabular}
 \end{ruledtabular}
 \caption{Chain and Primitive Path dimensions for PPA and Z1 as well as number of kinks $\ave{Z}$ for Z1
 for the LJ + FENE polymer melt. All
quantities given in reduced LJ units. It is remarkable that values
obtained via Z1 and PPA are very comparable, suggesting that chain
thickness and slippage effects seem to cancel as discussed in
\cite{shanbhag}.}
 \label{tab:chaindims}
\end{table*}
}


\newcommand{\TABLEABCDtwo}{
\begin{table*}[bthp]
\begin{ruledtabular}
\begin{tabular}{cccccccccc}
 & $l_0$ & $C(\infty)$ & $D$ & $Y$ & $A$ & $B$ & $G$ & $H$\\
system & &
        & \multicolumn{2}{c}{cf. Eq.\ (\ref{eq:reesqNG1})}
        & \multicolumn{2}{c}{cf. Eq.\ (\ref{eq:LppsqNG1})}
        & \multicolumn{2}{c}{cf. Eq.\ (\ref{Zepsilon1})} \\
\hline
LJ + FENE & $0.964$ & $1.852$ & $1.72$ & $3.55$ & $0.020$ & $1.04$ & $0.020$ & $0.12$ \\
polyethylene 450 K & $1.54$ \AA & $8.318$ & $19.7$ \AA$^2$& $131.4$ \AA$^2$& $0.22$ \AA$^2$ & $8.58$ \AA$^2$ &$0.023$ & $0.20$ \\
polyethylene 400 K & $1.54$ \AA & $8.535$ & $20.2$ \AA$^2$ & $85.3$ \AA$^2$& $0.24$ \AA$^2$ & $9.37$ \AA$^2$ &$0.025$ & $0.19$ \\
\end{tabular}
\caption{Data obtained via Z1. The coefficients $D$, $Y$, $A$, and
$B$ have been obtained from a least square fit to the available data
(covering $N\gg \Ne$) for $\ave{\Ree^2}$ and $\ave{\Lpp}$, according
to Eqs.\ (\ref{eq:reesqNG1}) and (\ref{eq:LppsqNG1}). Similarly,
coefficients $G$ and $H$ derive from the measured $\ave{Z}$ via
(\ref{Zepsilon1}).} \label{tabDYAC2}
\end{ruledtabular}
\end{table*}
}


\newcommand{\TABLENeNe}{
\begin{table*}[bthp]
\begin{ruledtabular}
\begin{tabular}{cccccccccccccc}
       & $\Ne$ & $\Nee(\Ne)$ & $\Ne$  & $\Nee(\Ne)$ & $\Ne$  & $\Nee(\Ne)$ & $\Ne$  & $\Nee(\Ne)$ & $\Ne$  & $\Nee(\Ne)$\\
       & \multicolumn{2}{c}{M--coil}
       & \multicolumn{2}{c}{approximate M--coil}
       & \multicolumn{2}{c}{simplified M--coil}
       & \multicolumn{2}{c}{M--kink}
       & \multicolumn{2}{c}{approximate M--kink}\\
system & \multicolumn{2}{c}{Eq.\ (\ref{Neestimatorcoil1})        }
       & \multicolumn{2}{c}{Eq.\ (\ref{eq:expNeestimatorcoil2})  }
       & \multicolumn{2}{c}{Eq.\ (\ref{eq:expconvstdEv})}
       & \multicolumn{2}{c}{Eq.\ (\ref{Neestimatorkink})        }
       & \multicolumn{2}{c}{Eq.\ (\ref{NelinearZ})        }
       \\
\hline
LJ + FENE &          $86.1$ & $87.8$ & $85.1$ & $89.6$ & $86.2$ & $90.1$  & $48.9$ & $46.3$ & $48.5$ & $55.7$\\
polyethylene 450 K & $84.0$ & $83.4$ & $84.4$ & $84.5$ & $90.6$ & $90.1$  & $44.2$ & $42.2$ & $43.3$ & $38.8$ \\
polyethylene 400 K & $82.3$ & $80.1$ & $80.5$ & $77.8$ & $83.9$ & $84.1$  & $41.5$ & $38.5$ & $40.1$ & $36.3$\\
\end{tabular}
\end{ruledtabular}
\caption{Data obtained via Z1. Selected results for $\Nee(N)$ for
all near-ideal M--coil and M--kink estimators defined in this
manuscript. For each estimator, two characteristic values are shown: $\Ne$
uses all available $N$ (up to $N=3500$ and $N=1000$ for the
LJ+FENE and PE models, respectively), and  $\Nee(\Ne)$ uses only data from short chains with $N\le \Ne$ (cf.\ Tab.\
\ref{tab:chaindims}). Values of $\Nee(\Ne)$ are thus obtained at
moderate computational cost, and are all in overall agreement with
$\Ne$. Approximate M--coil (M--kink) results should coincide with
M--coil (M--kink) results, if the relationships (\ref{eq:reesqNG1}),
(\ref{eq:LppsqNG1}) and (\ref{Zepsilon1}), respectively, accurately
hold. The simplified M--coil does not take into account the effect
of $C(N)$. M--coil (M--kink) is the estimator with the least
assumptions involved, if $\Ne$ needs to be estimated from coil
(kink) information (see also Appendix\ \ref{app:techMcoil}). The
fact that for all these estimators $\Nee(\Ne)\approx \Ne$ gives
sufficient evidence that these are in fact ideal estimators, in
sharp contrast to most S--estimators, quantitatively discussed in
Tab.\ \ref{tabNeNeearlier}.  Note that the very similar values of
$N_e$ reported for LJ+FENE and PE systems are a pure coincidence
arising from their similar values of $D/A$ (Tab.\ \ref{tabDYAC2};
cf.\ Eq.\ \ref{eq:expconvstdEv})} \label{tabNeNe}
\end{table*}
}


\newcommand{\TABLEEARLIERNeNe}{
\begin{table*}[bthp]
\begin{ruledtabular}
\begin{tabular}{cccccccccccccc}
       & $\Ne$ & $\Nee(\Ne)$ & $\Ne$ & $\Nee(\Ne)$ & $\Ne$ & $\Nee(\Ne)$ & $\Ne$ & $\Nee(\Ne)$\\
 & \multicolumn{2}{c}{classical S--coil}
       & \multicolumn{2}{c}{modified  S--coil}
       & \multicolumn{2}{c}{classical S--kink}
       & \multicolumn{2}{c}{modified  S--kink}        \\
system & \multicolumn{2}{c}{Eq.\ (\ref{eq:stdEveraers})}
       &\multicolumn{2}{c}{Eq.\ (\ref{eq:correctedEveraers})}
       & \multicolumn{2}{c}{Eq.\ (\ref{eq:stdKroger})}
       & \multicolumn{2}{c}{Eq.\ (\ref{eq:correctedKroger})}         \\
\hline
LJ + FENE &          $86.1$ & $40.0$ & $86.1$ & $129.7$ & $48.9$ & $31.7$ & $48.9$ & $51.2$ \\
polyethylene 450 K & $84.0$ & $39.7$ & $84.0$ & $192.8$ & $44.2$ & $30.4$ & $44.2$ & $48.3$ \\
polyethylene 400 K & $82.3$ & $37.3$ & $82.3$ & $191.8$ & $41.5$ & $28.8$ & $41.5$ & $44.9$ \\
\end{tabular}
\end{ruledtabular}
\caption{Data obtained via Z1. For comparison with Tab.\
\ref{tabNeNe}. Performance of previous S--coil and S--kink
estimators. Accurate $\Ne$--values have been overtaken from M--coil
and M--kink in Tab.\ \ref{tabNeNe}. Obviously, $\Nee(\Ne)$ is far
from being close to $\Ne$ in all cases, while the deviations are
strongest for the $\Ne$--estimates based on coils; the two kink
measures seem to at least bracket the true $\Ne$ (for the deeper
reason that $Z_0$, introduced in Sec.\ \ref{subsec:kinks}, must obey
$Z_0\in[-1,0]$).} \label{tabNeNeearlier}
\end{table*}
}


\title{Topological analysis of polymeric melts: Chain length effects and fast-converging estimators for
entanglement length}

\author{Robert S. Hoy} \email[Corresponding author. E--mail: ]{robertscotthoy@gmail.com}
\affiliation{Materials Research Laboratory, University of
California, Santa Barbara, CA 93106, U.S.A.}

\author{Katerina Foteinopoulou}
\affiliation{Institute for Optoelectronics and Microsystems (ISOM)
and ETSII, Universidad Polit\'ecnica de Madrid (UPM), Jos\'e
Guti\'errez Abascal 2, E--28006 Madrid, Spain}

\author{Martin Kr\"oger} \email[Electronic address: ]{http://www.complexfluids.ethz.ch}
\affiliation{Polymer Physics, ETH Z\"urich, Department of Materials,
CH--8093 Z\"urich, Switzerland}

\date{June 21, 2009}


\begin{abstract}
Primitive path analyses of entanglements are performed over a wide
range of chain lengths for both bead spring and atomistic
polyethylene polymer melts. Estimators for the entanglement length
$\Ne$ which operate on results for a single chain length $N$ are
shown to produce systematic $O(1/N)$ errors. The mathematical roots
of these errors are identified as (a) treating chain ends as
entanglements and (b) neglecting non-Gaussian corrections to chain
and primitive path dimensions. The prefactors for the $O(1/N)$
errors may be large; in general their magnitude depends both on the
polymer model and the method used to obtain primitive paths. We
propose, derive and test new estimators which eliminate these
systematic errors using information obtainable from the variation of
entanglement characteristics with chain length. The new estimators
produce accurate results for $\Ne$ from marginally entangled
systems. Formulas based on direct enumeration of entanglements
appear to converge faster and are simpler to apply.
\end{abstract}

\maketitle

\section{Introduction}
 \label{sec:intro}

The features of polymer melt rheology are determined primarily by
the random-walk-like structure of the constituent chains and the
fact that chains cannot cross. The motion of sufficiently long
chains is limited by ``entanglements" which are topological
constraints imposed by the other chains. These become important and
dramatically change many melt properties (e.\ g.\ diffusivity and
viscosity) as the degree of polymerization becomes larger than the
``entanglement length" $\Ne$. The value of $\Ne$ is both a key
quantity measured in mechanical and rheological experiments and a
key parameter in tube theories of dense polymeric systems
\cite{doibook}.

$\Ne$ is usually considered to be a number set by chemistry and
thermodynamic conditions (e.\ g.\ chain stiffness, concentration,
and temperature). It has been empirically related to a ``packing''
length \cite{packing}; $\Ne\propto (\rho b^3)^{-2}$ \cite{kh2000},
where $\rho$ is monomer number density and $b^2=\ave{\Ree^2/(N-1)}$
is the statistical segment length of chains with end-to-end distance
$\Ree$ and mean degree of polymerization $N$. In terms of individual
entanglements, $\Ne$ is defined as the ratio between $N$ and the
mean number of entanglements per chain $\ZZ$, in the limit of
infinite chain length, \be
 \Ne = \lim_{N\rightarrow \infty} \frac{N}{\ZZ}. \label{Nedef}
\ee
We call a function $\Nee(N)$ an $\Ne$--estimate if it has the
property
\be \lim_{N\rightarrow \infty} \Nee(N) = \Ne,
\label{Nerequire}
\ee
where $\Ne$ is a system dependent but $N$-independent quantity.
Comparing Eq.\ (\ref{Nedef}) with (\ref{Nerequire}) does \textit{not} imply choosing $\Nee(N)=N/\ZZ$.
The typical experimental $\Ne$--estimate uses the plateau modulus $G_{N}^{0}$ \cite{doibook}:
\be
\Nee(N) = \displaystyle\frac{4m\rho k_\textrm{B} T}{5G_{N}^{0}},
\label{eq:experimentalestimator}
\ee
where $m$ is monomer mass,  $k_\textrm{B}$ is the Boltzmann constant, and $T$ is temperature.

A closely related theoretical construct is the primitive path (PP),
defined by Edwards \cite{edwards77} as the shortest path a chain
fixed at its ends can follow without crossing any other chains.
Rubinstein and Helfand \cite{rubinstein85} realized that the
entanglement network of a system could be obtained by reducing all
chains to their PPs simultaneously. Such a reduction process is
analytically intractable, but has recently been achieved by computer
simulations
\cite{everscience,sukumaranlong,Zhou05,cpc,christos,shanbhag06,Foteinopoulou2006,jammedHS}
which generate networks of PPs from model polymer melts, glasses,
random jammed packings and solutions. These simulations estimate
$\Ne$ either from the chain statistics of the PPs
\cite{everscience,sukumaranlong,Zhou05} or from direct enumeration
of entanglements (contacts between PPs)
\cite{cpc,christos,shanbhag,shanbhag06,Foteinopoulou2006,christoscurropin}, which determines
$\ZZ$.

Chain-statistical and direct enumeration approaches produce
different results for $\Ne$ for the same atomistic configurations,
suggesting that ``rheological'' and ``topological'' entanglements
are not equivalent \cite{christoscurropin}. This discrepancy has
been attributed to the fact that chemical distances between
entanglements are not uniform, but rather are drawn from broad
distributions
\cite{mkbook,cpc,christoscurropin,Foteinopoulou2006,shanbhag}, even
at equilibrium. Studies of how entanglement properties change with
$N$ are therefore of obvious interest. Moreover, primitive path
statistics enter recently developed sliplink--based models
\cite{sliplink}.

In this paper we seek an ``ideal'' $\Ne$-estimate which approaches
$\Ne$ at the smallest possible $N$. There have been several attempts
in the literature, summarized below, to derive $\Ne$-estimates, but
these have all exhibited poor convergence (i.\ e.\ by approaching
$N_e$ only at large $N\gg \Ne$). Molecular dynamics simulation times
increase with chain length $N$ approximately as $N^5$ at large $N$
(relaxation time $\tau\propto N^{3.5}$ times system size $\propto
N^{3/2}$), so improved $\Ne$--estimates have obvious benefits for
computationally efficient determination of $\Ne$. By analyzing a
number of coarse-grained and atomistic systems, we find a rather
general solution to this problem of setting up a $\Ne$--estimator
which allows to predict $\Ne$ from weakly entangled linear polymer
melts.

The organization of this paper is as follows. Section
\ref{sec:methodsdims} presents the polymer models used here and the
topological analysis methods which provide us with the entanglement
network (primitive paths). Section \ref{sec:prelim} distinguishes
between valid and quickly converging (ideal) $\Ne$--estimators, and
discusses some model-- and method--independent issues with existing
estimators. Examples are given which highlight systematic errors
caused by improper treatment of chain ends and of the non-Gaussian
statistics of chains and primitive paths. Section \ref{sec:ideal}
derives two (potentially) near--ideal estimators which extract $\Ne$
from the variation of entanglement characteristics with $N$. Section
\ref{sec:performance} presents and discusses numerical results for
these estimators for two very different model polymers. We verify
that they are basically ideal, explain why this is so, and derive
simplified forms which further illustrate the connection of $N_e$ to
chain structure and entanglement statistics and are also near-ideal.
Section \ref{sec:conclude} contains conclusions, and two Appendices
provide additional technical details.

\section{Polymer Models and Methods}
\label{sec:methodsdims}

\subsection{Polymer model systems}

We have created thoroughly equilibrated configurations for two very
different (but commonly used) model polymer melts;  monodisperse
`Kremer--Grest' bead--spring chains, and atomistic, polydisperse
polyethylene. These two are chosen because they have similar values
of $N_e$ but very different chain stiffness constants $C(\infty)$.
Polyethylene is much more ``tightly entangled'' \cite{uchidaa2008}
in the sense of having a much lower value of $N_e/C(\infty)$; cf.
Tabs.\ \ref{tabDYAC2} and  \ref{tabNeNe}.

The bead spring model \cite{kremer90} captures the features of
polymers which are key to entanglement physics, most importantly
chain connectivity/uncrossability. Each chain contains $N$ beads of
mass $m$. All beads interact via the truncated and shifted
Lennard-Jones potential $U_\textrm{LJ}(r) =
4\epsilon_\textrm{LJ}[(\sigma/r)^{12} - (\sigma/r)^{6} -
(\sigma/r_{c})^{12} + (\sigma/r_c)^{6}]$, where $r_{c} =
2^{1/6}\sigma$ is the cutoff radius and $U_\textrm{LJ}(r) = 0$ for
$r > r_{c}$. Here $\sigma$ is the bead diameter and
$\epsilon_\textrm{LJ}$ is the binding energy, which are both set to
$1$; all quantities will thus be dimensionless and given in the
conventional Lennard--Jones units. Covalent bonds between adjacent
monomers on a chain are modeled using the FENE potential $U(r) =
-\frac{1}{2} kR_{0}^2 \ln [1 - (r/R_{0})^{2}]$, with the canonical
parameter choices $R_{0} = 1.5$ and $k = 30$\ \cite{kremer90}. The
equilibrium bond length is $l_0\approx 0.96$. This model is
hereafter referred to as the ``LJ + FENE'' model.

Values of the density and temperature ($\rho = 0.85$ and $T = 1.0$)
are those typically used for melt simulations
\cite{kremer90,everscience,sukumaranlong}. All systems contain
280,000 total beads. While all are monodisperse, we employ a wide
range of chain lengths, $4 \leq N \leq 3500$. Those with $N \geq
100$ are equilibrated using the ``double-bridging hybrid'' (DBH)
algorithm \cite{everaers}. DBH uses molecular dynamics to update
monomer positions and Monte Carlo chain-topology-altering moves
\cite{endbridge3} to overcome the slow diffusive dynamics
\cite{doibook} of entangled chains. All equilibration simulations
were performed using the LAMMPS \cite{LAMMPS} molecular dynamics
code. Ref.\ \cite{kremer90} predicted $\Ne\approx 35$ at the
above--mentioned state point using various ``rheological'' measures
applied to systems with $N\le 400$, while a similar analysis in
\cite{putz00} predicted $\Ne\approx 75$.

In all simulations of the atomistic polyethylene (PE) melt, the
united atom (UA) representation is adopted. Accordingly, carbon
atoms along with their bonded hydrogens are lumped into single
spherical interacting sites. There is no distinction between methyl
and methylene units in the interaction potentials. All bond lengths
are kept constant ($l_0 = 1.54$ \AA), while bending and torsion
angles are respectively governed by harmonic and sum-of-cosines
potentials \cite{foteinopoulou08PE,kf23}. Pair interactions between
all intermolecular neighbors, and intramolecular neighbors separated
by more than three bonds, are described by the $12$--$6$
Lennard-Jones potential. The parameters of the mathematical formulas
for the bonded and non-bonded interactions are given in Refs.\
\cite{Foteinopoulou2006,foteinopoulou08PE,endbridge3,kf23}. These
interaction potentials yield accurate predictions of the volumetric,
structural and conformational properties of PE melts over a wide
range of chain lengths and temperatures
\cite{foteinopoulou08PE,endbridge3}.

All atomistic PE systems were equilibrated through Monte Carlo (MC)
simulations based on advanced chain-connectivity-altering
algorithms: the end-bridging \cite{Pant1995} and double bridging
\cite{endbridge3,kf23} moves along with their intramolecular
variants. The simulated systems are characterized by average chain
lengths from $N=24$ up to $N=1000$, with a small degree of
polydispersity. Chain lengths are uniformly distributed over the
interval $[(1-\triangle)N, (1+\triangle)N]$. Here $\triangle$, the
half width of the uniform chain length distribution reduced by $N$,
is $0.5$ and $0.4$ for $24\le N\le 224$ and $270\le N\le 1000$,
respectively. More details about the MC scheme, including a full
list of moves, attempt probabilities and acceptance rates, can be
found elsewhere \cite{foteinopoulou08PE}. Equilibration at all
length scales, which is essential to obtaining meaningful results
from entanglement analyses \cite{hoy2005}, was verified using
several metrics \cite{foteinopoulou08PE}. In this study, results are
presented for $T=400$ K and $T = 450$ K, both for $P = 1$ atm.

\subsection{Entanglement network and primitive paths}

\TABLECHAINDIMS

For the melt configurations the reduction to primitive paths was
performed using two methods, PPA and Z, using the procedures
described in Refs.\ \cite{everscience, sukumaranlong, cpc,shanbhag}.
PPA simulations used LAMMPS and Z simulations used the Z1 code
\cite{Z1online}.  Both PPA and Z1 analyses are performed for the LJ+FENE model, while only Z1 analysis is performed for PE. In both methods, all chain ends are fixed in
space. Intrachain excluded volume interactions are disabled while
chain uncrossability is retained. Both classical PPA
\cite{everscience} and geometrical methods (Z1 \cite{cpc,shanbhag}
or CReTA \cite{christos,christoscurropin}) provide the configuration
of the entanglement network and the contour lengths $\Lpp$ of each
primitive path. In PPA, disabling intrachain excluded volume
produces a tensile force \cite{elastic} in chains which reduces the
contour lengths. In Z1, contour lengths are monotonically reduced
through geometrical moves in the limit of zero primitive chain
thickness. In addition to $\Lpp$ and the configuration of the
entanglement network, $Z1$ analysis also yields the number of
interior ``kinks'' \cite{cpc}, $Z$, in the three-dimensional
primitive path of each chain. $\ZZ$ is considered to be proportional
to the number of entanglements, regardless of the details of the
definition used to define an entanglement.

Runs end when the mean length of the primitive paths, $\ave{\Lpp}$,
and/or the mean number of interior kinks per chain, $\ZZ$, converge.
Self entanglements are neglected, but their number is
inconsequential for the systems considered here
\cite{sukumaranlong}. The CReTA method works similarly, and the
conclusions reached here for Z1 analysis should apply similarly to
CReTA results \cite{christoscurropin,shanbhag}.

Table \ref{tab:chaindims} summarizes chain and primitive path
dimensions as well as $\ZZ$ for LJ+FENE chains with $20 \leq N \leq
3500$. Statistically independent initial states were used so that
the random error on all quantities is $\lesssim 2.5\%$. It is
remarkable that PPA and Z1 data for $\ave{\Lpp}$ and also
$\ave{\Lpp^2}$ are so similar, considering the differences between
the contour length reduction methods. Relative to Z1 results, PPA
values of $\ave{\Lpp}$ are increased by finite chain thickness
effects \cite{christos, hoy2007} and decreased by chain end slipoff
\cite{shanbhag}. Both these effects should decrease in strength as
$N$ increases, and indeed
$\ave{\Lpp^2}_\textrm{PPA}/\ave{\Lpp^2}_\textrm{Z1}$ decreases from
$\sim 1.17$ to $\sim 1.06$ over the range $20 \leq N \leq 3500$. A
very comparable trend is offered by
$\ave{\Lpp}^2_\textrm{PPA}/\ave{\Lpp}^2_\textrm{Z1}$.

PPA results for the shortest chains $(N < 20)$ are not presented.
Standard PPA is unreliable for very short chains because the
presence of a high concentration of fixed chain ends combined with
the finite bead diameter effectively inhibits relaxation
\cite{christos,hoy2007}. These problems are even worse for
topological analysis of lattice polymer systems, see e.\ g.\ Ref.\
\cite{shanbhag06}. In the following, where (as will be shown)
accurate data from very short chains is important, we focus on Z1
results.

\section{Towards Valid Estimators}
\label{sec:prelim}

A basic task of topological analysis is to calculate $\Ne$ from the
full microscopic configuration of the entanglement network. The
simplest approaches employ only the mean square end--to--end
distance of chains, $\ave{\Ree^2}$ and either the mean length of the
primitive paths, $\ave{\Lpp}$ or the mean number of kinks,
$\ave{Z}$. Notice that $\ZZ$ is not an integer, but semipositive,
$\ZZ\ge 0$. In order to estimate $\Ne$ from weakly entangled systems
one of course needs physical insight; when this is limited, a good
$\Ne$--estimator can only be guessed.

Some restrictions arise from a purely mathematical viewpoint. A
valid estimator $\Nee(N)$ has the following properties.
It\begin{itemize}
\item[\BBlue{(i)}] obeys Eq.\ (\ref{Nerequire}) and uses information from polymer configurations
whose mean chain length does not exceed $N$;
\item[\BBlue{(ii)}] either yields $\Nee(N)\ge N$
or leaves $\Nee(N)$ undefined for a system of completely unentangled ($\ave{Z}=0$)
chains.
\end{itemize}
An ``ideal'' estimator we define to
\begin{itemize}
\item[\BBlue{(iii)}] correctly predict
$\Ne$ for all $N$ exceeding $\Ne$, or for all $\ZZ$ exceeding unity.
\end{itemize}
Accordingly, for an ideal estimator, the following weaker conditions
hold. An ideal estimator
\begin{itemize}
\item[\BBlue{(iv)}] diverges for a
system of rodlike chains possessing $\Ne=\infty$, and
\item[\BBlue{(v)}]
exhibits $\Nee(N)\le N$ when each chain has in average more than a
single entanglement, $\ZZ> 1$.
\end{itemize}

The following two subsections repeat earlier approaches to estimate
$\Ne$. Basic considerations of finite chain length effects, errors
from improper treatment of non-Gaussian structure, and the general
behavior of quantities entering $\Ne$ are discussed. These
subsections are meant to prepare the reader for the ideal estimators
to be presented in Sec.\ \ref{sec:ideal}. They reflect the
chronology of our search for better estimators and help the reader
to understand the magnitude of improvements presented in Section
\ref{sec:performance}. The arguments given here ultimately point the
way to construct ideal estimators.

\subsection{Non-Ideal Estimators}

Modelling primitive paths as random walks, Everaers \textit{et.\
al.} \cite{everscience} developed an estimator (which we denote as
"classical S--coil") which operates on results for configurations
("coils") of a single (S) chain length,
\begin{equation}
\Nee(N)=(N-1)\displaystyle\frac{\ave{\Ree^2}}{\ave{\Lpp}^2}.
\label{eq:stdEveraers}
\end{equation}
The classical S--coil estimate (\ref{eq:stdEveraers}) is useful
because (for long chains) it relates changes in chain structure to
rheological trends \cite{everscience,uchidaa2008}. However, while it
fulfills basic requirements (i) and (ii) (both unentangled and
rodlike chains have $\Ree=\Lpp$), it lacks properties (iii) and
(iv). As the exact relaction of $\ave{\Ree^2}/\ave{\Lpp}^2$ and
$\ZZ$ is unknown, it is {\it a priori} unclear wether it has
property (v).

The corresponding estimator operating on the number of kinks, $\ZZ$,
and originally employed in \cite{cpc}, denoted here as ``classical S--kink'', is
\begin{equation}
\Nee(N) = \displaystyle\frac{N(N-1)}{\ZZ(N-1)+N},
\label{eq:stdKroger}
\end{equation}
which fulfills the basic requirements (i) and (ii), and also (v),
but lacks (iii) according to Ref.\ \cite{Foteinopoulou2006} and
(iv) by definition. The presence of both $N-1$ and $N$ in Eqs.\
(\ref{eq:stdEveraers}), (\ref{eq:stdKroger}), and subsequent
estimators reflects the fact that it is the existence of a bond
rather than a bead which is responsible for the presence or absence
of an entanglement between two chain contours.

\begin{figure}[tbh]
\includegraphics[width=7.5cm]{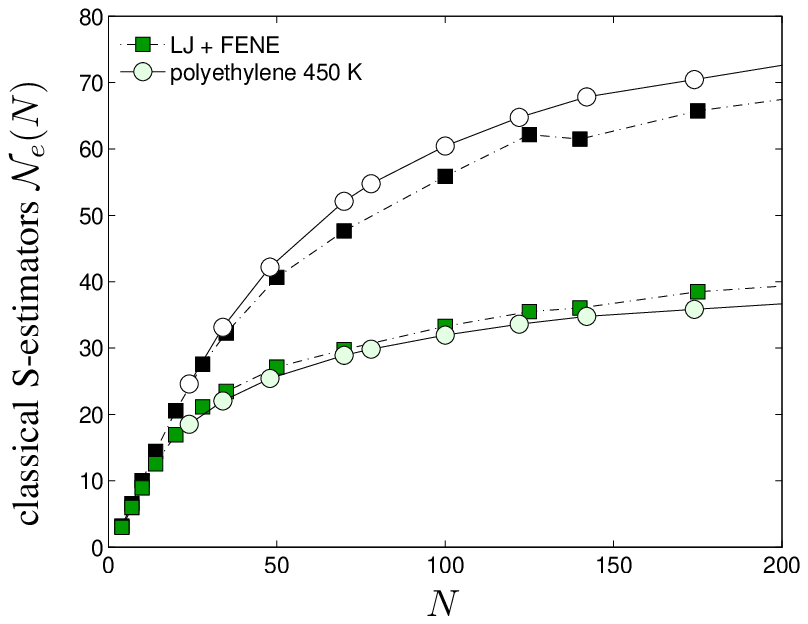}
 \caption{(color online)
Performance of classical $\Ne$--estimators $\Nee(N)$ based on coils
(\ref{eq:stdEveraers}) (upper two curves) and kinks
(\ref{eq:stdKroger}). Data is shown for the two model polymer melts
studied in this manuscript.  The trends with $N$ is in agreement
with published results for other systems
\cite{cpc,Foteinopoulou2006,christos,shanbhag,leon05,
subramanian08,harmandaris09}. The convergence behavior is poor, as
$\Ne\equiv\lim_{N\rightarrow\infty}\Nee(N)$ obviously cannot be
extrapolated studying chains with $N < 100$, while $\Ne$ turns out
to stay well below $100$ for both systems. An "ideal estimator", as
defined in Sec.~\ref{sec:prelim}, would converge when $N$ exceeds
$\Ne$ or earlier.\label{Fig_old_standard_estimators}}
\end{figure}

The performance of the two classical estimators
(\ref{eq:stdEveraers}) and (\ref{eq:stdKroger}) for the two polymer
models considered here is illustrated in Fig.\
\ref{Fig_old_standard_estimators}. Values of $\Nee(N)$ converge very
slowly with increasing $N$. As expected from their form, but
contrary to both rheological intuition and condition (ii) , values
of $\Nee(N)$ drop strongly with decreasing $N$. For marginally
entangled chains (where $N$ is just large enough so that $\ZZ$ is
small but nonzero), both classical estimators yield $\Nee(N)\le
N-1$. For example, for $N = 20$, they both predict $\Nee(N) = 17$,
which is close to the (improper) upper bound $N-1=19$. This
prediction obviously has no connection to the actual topology of the
system.

Thus Eqs.\ (\ref{eq:stdEveraers}) and (\ref{eq:stdKroger}) always
underestimate, but never overestimate $\Ne$.
This feature of the two estimators in the limit of unentangled chains is
particularly (if retrospectively) disappointing, as it is
incompatible with goal (iii). Similar behavior was reported (but not
analyzed as in this paper) in Refs.\ \cite{leon05,
subramanian08,harmandaris09}.

Other previously published $\Ne$--estimators
\cite{sukumaranlong,christos,mkbook,moorthi} also have some, but not
all, of properties (i)-(v). One of the most promising was proposed
in Ref.\ \cite{sukumaranlong}. It estimates $\Ne$ from the
\textit{internal} statistics of primitive paths, for a single $N$.
The squared Euclidean distances $\ave{R^2(n)}$ between monomers
separated by chemical distance $n\le N-1$ \textit{after} topological
analysis (i.\ e., the chain statistics of the primitive paths) were
fit \cite{sukumaranlong} to those of a freely rotating chain with
fixed bond length fixed bending angle. $\Ne$ was then identified
with the chain stiffness constant $C(\infty)$ of the freely rotating
chain \cite{footCN}. This estimator does not \textit{obviously} fail
to meet any of conditions (i)-(v). In Ref.\ \cite{sukumaranlong} it
gave values of $\Nee(N)$ which decreased more slowly than Eq.\
(\ref{eq:stdEveraers}) as $N$ decreased. Unfortunately, its
predictions agree with Eq.\ (\ref{eq:stdEveraers}) at moderate $N
\gtrsim 100$ and thus it fails condition (iii).

New S-estimators based on modifications to Eqs.\
(\ref{eq:stdEveraers}) and (\ref{eq:stdKroger}) may be proposed.
During the course of developing ideal estimators (to be introduced
in Sec.\ \ref{sec:ideal}), we developed two modified single chain
length estimators which tend to approach $\Ne$ from above rather
from below. These are the "modified S--kink" estimator
\begin{equation}
\Nee(N)=\displaystyle\frac{N}{\ZZ},
\label{eq:correctedKroger}
\end{equation}
and the mathematically similar "modified S--coil" estimator
\begin{equation}
\Nee(N)=(N-1)\left(\displaystyle\frac{\ave{\Lpp^2}}{\ave{\Ree^2}}-1\right)^{-1}.
\label{eq:correctedEveraers}
\end{equation}
A motivation for the use of  $\ave{\Lpp^2}$ rather than
$\ave{\Lpp}^2$ in Eq.\ (\ref{eq:correctedEveraers}) appears in
Appendix \ref{app:fluctuations}. Fig.\
\ref{Fig_old_corrected_estimators} shows results for Eqs.\
(\ref{eq:correctedKroger}) and (\ref{eq:correctedEveraers}) for the
same systems analyzed in Fig.\ \ref{Fig_old_standard_estimators}.
Both modified single--chain estimators give $\Nee(N)=\infty$ for
unentangled chains, thus fulfilling criterion (iv) in addition to
(i) and (ii), but they still fail to fulfill goal (iii) since they
tend to overestimate $\Ne$ for weakly entangled chains.

\begin{figure}[tbh]
\includegraphics[width=7.5cm]{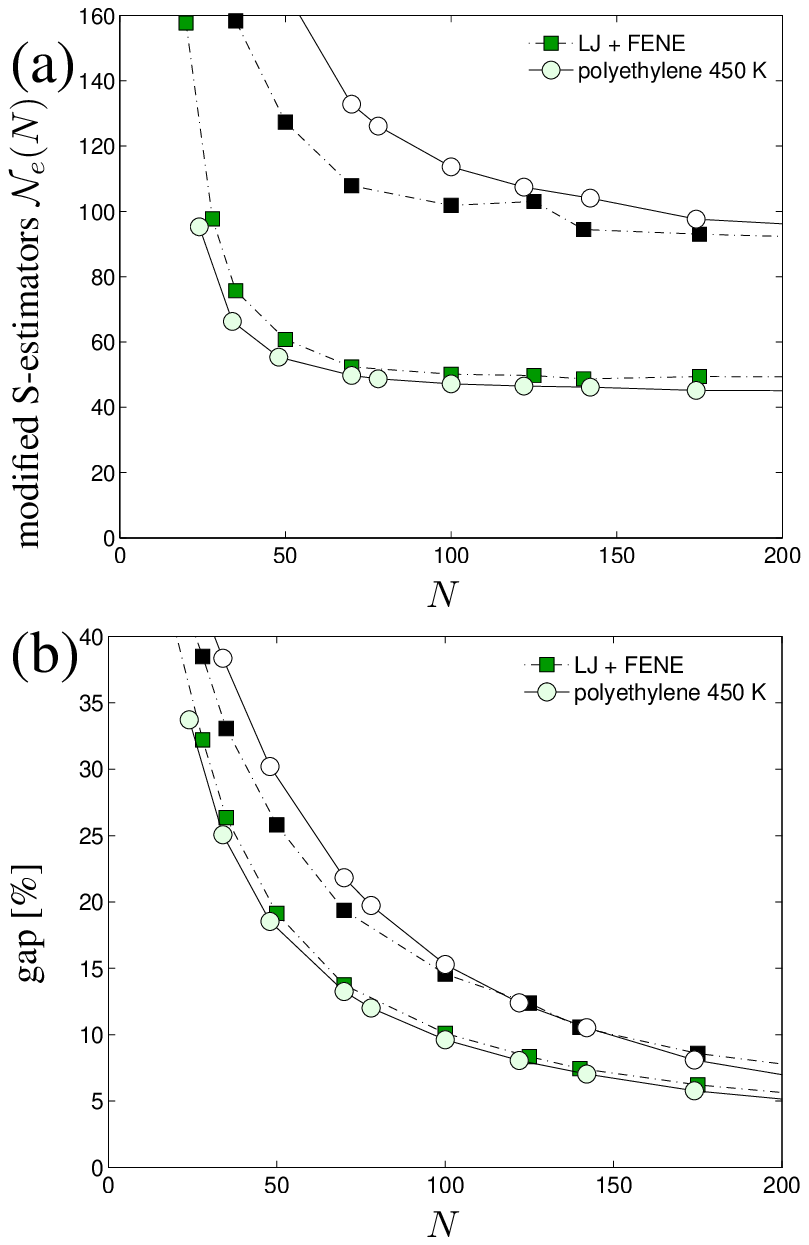}
 \caption{(color online) (a)
Performance of modified S--kink estimator (\ref{eq:correctedKroger})
(lower two curves) and the modified S--coil
(\ref{eq:correctedEveraers}) (upper two curves). which approach
$\Ne$ from above. Data are for the same systems analyzed in Fig.\
\ref{Fig_old_standard_estimators}. The single-configuration
estimator for kinks exhibits an improved convergence behavior
compared with (\ref{eq:stdKroger}). Under circumstances discussed in
Sec.\ \ref{sec:performance}, application of both modified and
original classical estimators allows one to obtain lower and upper
bounds on $\Ne$ which tighten with increasing $N$.  (b) Shown are
the relative differences (``gap [\%]'') between $\Nee(N)$ values
shown in Fig.~\ref{Fig_old_standard_estimators} and the ones plotted
in part (a) of the current graph. Differences are smaller for $\Ne$
estimated from kinks (lower two
curves).\label{Fig_old_corrected_estimators}}
\end{figure}

\subsection{Errors from improper treatment of non-Gaussian structure and chain ends}

Critically,  none of the above-mentioned estimators seem to be able
to predict $\Nee(N)=\Ne$ for weakly entangled systems with a
slightly positive $\ZZ\lesssim 1$. All above--cited previous works
as well as Eq.\ (\ref{eq:correctedEveraers}) have only produced
convergence for $N \gg \Ne$, and we are not aware of any studies
where convergence has been achieved at $N\approx \Ne$, i.\ e., we
are not aware of the former existence of any ideal $\Ne$--estimator.
However, the failure of so many previous attempts both makes it
worth examining the common reasons why they have failed, and in fact
points the way to creating ideal $\Ne$--estimators.

\begin{figure}[tbh]
\includegraphics[width=7.5cm]{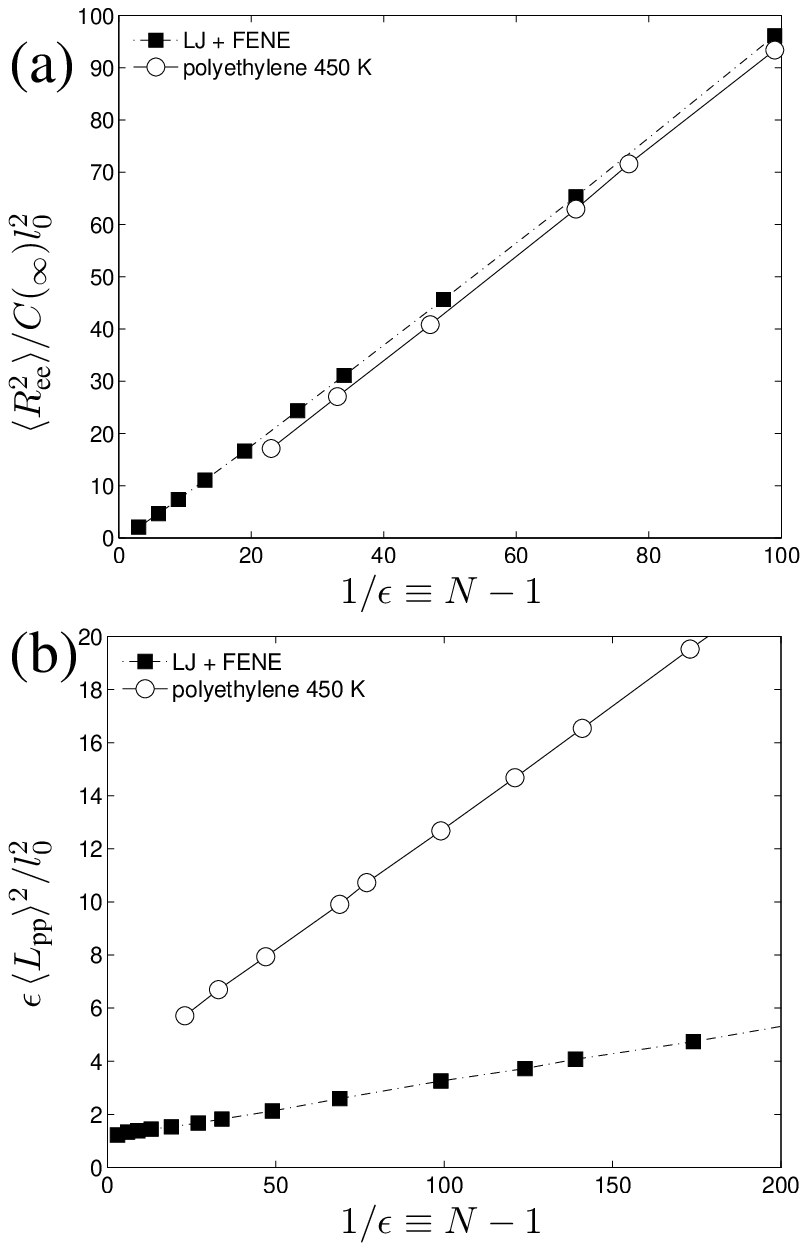}
 \caption{(a) Testing the
applicability of Eq.\ (\ref{eq:reesqNG1}) which predicts linear
behavior (slope $\propto D$, offset $\propto Y$) in this
representation. We obtain $C(\infty) \approx 1.85$ and $C(\infty)
\approx 8.3$ for the LJ+FENE and PE models, respectively (cf. Tab.\
\ref{tabDYAC2}). (b) Testing the validity of Eq.\
(\ref{eq:LppsqNG1}) for both types of melts (slope $\propto A$,
offset $\propto B$). The linear relationship is employed to derive
estimator (\ref{Neestimatorcoil1}) in Sec.\ \ref{subsec:coils}.
(a+b) Data for larger $N$ are not shown but also agree to all
displayed fit lines, to within statistical
errors.\label{Fig_Ree2_div_b02_Cinfty}}
\end{figure}

To leading order in $\epsilon\equiv (N-1)^{-1}$ (i.\ e.\ the inverse
number of bonds), data for a wide variety of model polymers (see e.\
g.\ Refs.\ \cite{everaers, harmandaris09,foteinopoulou08PE}), as
well as the data obtained in this study (see Fig.\
\ref{Fig_Ree2_div_b02_Cinfty}a) are consistent with
 \begin{equation}
 \ave{\Ree^2}(\epsilon) = D/\epsilon - Y,
 \label{eq:reesqNG1}
 \end{equation}
where the relative magnitudes of the constant coefficients $Y$ and
$D$ depend on factors such as chain stiffness, molecular details and
thermodynamic conditions.

Also, orientations of successive PP segments are
correlated \cite{christos}, so $\ave{\Lpp}^2$ should not be simply
quadratic in chain length. The expected leading order behavior of
$\ave{\Lpp}^2$ is
 \bea
 \ave{\Lpp}^2(\epsilon) = A/\epsilon^{2} + B/\epsilon,
 \label{eq:LppsqNG1}
 \eea
where $B$ contains contributions from non-Gaussian statistics and
contour length fluctuations \cite{doibook}. Relationships\
(\ref{eq:reesqNG1}) and (\ref{eq:LppsqNG1}) are consistent with data
reported elsewhere (e.\ g.\ Refs.\
\cite{subramanian08,foteinopoulou08PE}) as well with our own data,
as shown in Fig.\ \ref{Fig_Ree2_div_b02_Cinfty}.

At this point it is worthwhile to mention that we are going to make
use of (\ref{eq:LppsqNG1}), which is able to capture our results for
$\ave{L_{pp}^2}$ down to chain lengths $N$ small compared with
$\Ne$, to devise an ideal estimator in Sec.\ \ref{sec:ideal}.
Relationship (\ref{eq:reesqNG1}) however, as we will see, will
\textit{not} be required to hold to devise an ideal estimator.

Inserting Eqs.~(\ref{eq:reesqNG1}) and (\ref{eq:LppsqNG1}) into the
classical and modified S--coil Eqs.\ (\ref{eq:stdEveraers}) and (\ref{eq:correctedEveraers})
respectively give, to leading order in
$\epsilon$,
\beas{ecoils1}
 \Nee(N) &\stackrel{(\ref{eq:stdEveraers})}{=}& \displaystyle\frac{D}{A} - \displaystyle\frac{AY +
 BD}{A^2}\,\epsilon + O(\epsilon^2), \label{PREeq:expconvstdEv} \\
 \Nee(N) &\stackrel{(\ref{eq:correctedEveraers})}{=}& \displaystyle\frac{D}{A} + \displaystyle\frac{D^2 - AY - BD}{A^2}\,\epsilon + O(\epsilon^2), \label{eq:expconvcorrEv}
  \label{eq:expNeestimatorcoil}
\eeas Thus non-Gaussian structure of both chains and primitive paths
naturally lead to systematic $O(\epsilon) \simeq O(1/N)$ errors in
earlier estimators for $\Ne$ \cite{alexpriv}.

Similarly, $\ZZ$ necessarily scales as $\epsilon^{-1}$ in the
$N\to\infty$ limit. In the same spirit as the above analysis, and
noting the failure (Fig.\ \ref{Fig_old_corrected_estimators}) of
Eqs.\ (\ref{eq:stdKroger}) and (\ref{eq:correctedKroger}) to meet
condition (iii), let us hypothesize that  finite chain length leads
to the leading order behavior \be
 \ave{Z}(\epsilon) = G/\epsilon - H,
 \label{Zepsilon1}
\ee where $G$ and $H$ are both positive. This assumption is actually
consistent with the data in Tab.\ \ref{tab:chaindims} and previous
works \cite{foteinopoulou08PE}; see also Sec.\
\ref{sec:ideal}. The classical and modified S--kink Eqs.
(\ref{eq:stdKroger}) and (\ref{eq:correctedKroger}) then become
\beas{ekinks1}
 \Nee(N) &\stackrel{(\ref{Zepsilon1})\;\textrm{in}\;(\ref{eq:stdKroger})}{=}& \frac{1}{G} - \frac{1-G-H}{G^2}\,\epsilon + O(\epsilon^2), \\
 \Nee(N) &\stackrel{(\ref{Zepsilon1})\;\textrm{in}\;(\ref{eq:correctedKroger})}{=}& \frac{1}{G} + \frac{G+H}{G^2}\,\epsilon + O(\epsilon^2).
\eeas Again, systematic $O(\epsilon)$ errors are predicted. In this
case, however, the source is chains being too short  to be in the
asymptotic entangled limit defined by Eq.\ (\ref{Nedef}).

A key to understanding the failure of previous $\Ne$--estimators is that \textit{differences} in the prefactors of the $O(\epsilon)$ errors [Eqs.\ (\ref{ecoils1}) and (\ref{ekinks1})] arise from different treatment of chain ends.
The classical S--kink equation (\ref{eq:stdKroger}) underestimates $\Ne$ as long as $G+H<1$,
and the modified S--kink equation (\ref{eq:correctedKroger}) strictly overestimates $\Ne$, since both $G$ and $H$ are positive.
Similarly, the prefactor $(AY + BD - D^2)/A^{2}$
(Eq.\ \ref{eq:expconvcorrEv}) of the systematic $O(\epsilon)$ error
in the modified S--coil equation (\ref{eq:correctedEveraers}) contains two contributions of
different origins.
$(AY + BD)/A^2$ arises from the Gaussian-chain approximation used, while $-D^2/A^2$ arises from the attempt to correct for chain ends effects (i.\ e.\ the ``-1'').

\TABLEABCDtwo

We have determined the coefficients $A$, $B$, $D$, $G$, $H$, and $Y$
using all available data from our simulations; their values for both
polymer models are shown in Tab.\ \ref{tabDYAC2}. Coincidentally,
for LJ+FENE chains, $(AY+BD)/A^2 \simeq 5 \times10^3$ and $D^2/A^2
\simeq 7 \times 10^3$. The systematic $O(\epsilon)$ error for the
modified S--coil (\ref{eq:correctedEveraers}) is actually small for
LJ+FENE systems due to the near-cancellation of its contributing
terms. There is no reason to believe this behavior is general, and
tests on additional polymer models would be necessary
\cite{alexpriv} to better characterize how rapidly the modified
S--coil typically converges. However, it is reasonable to expect it
typically converges more rapidly than the classical S--coil
(\ref{eq:stdEveraers}).

Before turning to ideal estimators, we mention that the modified
S--kink (\ref{eq:correctedKroger}) can be regarded as corrected
version of classical S--kink (\ref{eq:stdKroger}), as it eliminates
an $O(\epsilon)$ error from the latter, and thus converges faster.

\section{Ideal estimators}
\label{sec:ideal}

Given the prevalence of subtle systematic $O(\epsilon)$ errors in
non-ideal $\Ne$--estimators, it is reasonable to suppose that in
developing an ideal estimator, one has the freedom to introduce
system-dependent (but $N$--independent) coefficients, e.\ g., $c$,
$c'$, and $Z_0$, in equations for a valid $\Nee(N)$ such as $N/\ZZ +
c\,\epsilon$, $N(1-c'\epsilon)/\ZZ$, or $N/(\ZZ+Z_0)$. These
formulae are all potentially valid estimators because they fulfill
the basic requirement (Eq.\ \ref{Nerequire}). The coefficients are
somewhat related to each other, but have slightly different physical
meanings. They fulfill conditions (i) and (ii) for arbitrary $c$ and
$c'$, but only if $Z_0\le 1$. Note that finding an ideal
$\Ne$--estimator neither depends on the interpretation of $\ZZ$ or
requires \textit{a priori} knowledge of the numerical values of the
coefficients.  However, these numerical values are required to turn
the above three expressions into $\Ne$--estimates before they can be
applied. As these numerical values are certainly sensitive to system
features like chain thickness and stiffness, it is impossible to
determine them from a single set of $\ave{\Lpp}$, $\ave{\Ree^2}$,
and $\ZZ$ values.

The best possible estimator gives $\Nee(N)=\Ne$ for $\ZZ\ll 1$, but
such an estimator would have to rely on incomplete information, some
model assumptions, or make use of some `universal' features of
entangled systems such as those suggested by Refs.\
\cite{Foteinopoulou2006,christos}. We make use of two such findings (Sec.\ \ref{sec:prelim}):
for the polymer models considered here, both $\ZZ$ and
$\ave{\Lpp}^2/(N-1)$ are linear in $N$ above certain characteristic
thresholds. Further supporting data for atomistic polyethylene have been
reported recently by some of us \cite{foteinopoulou08PE}. 

For both models considered here, the `characteristic thresholds' are located at $\ZZ<1$ and $N<\Ne$, allowing us to make use of the `linearities' to construct ideal $\Ne$--estimators. 
We now derive two near-ideal $\Ne$--estimators, for kinks and coils
respectively. These estimators operate on multiple (M) systems with
different chain lengths, rather than on a single configuration, and
will be denoted as M-coil and M-kink in order to clearly distinguish
between S-- and M--estimators.
Careful empirical tests of the new estimators' validity is quite essential,
and will be given in Sec.\ \ref{sec:performance}.

Below, the idea behind the different roles of Eqs.\
(\ref{eq:reesqNG1}), (\ref{eq:LppsqNG1}), and (\ref{Zepsilon1}) is that the statistics
of the entanglement network can be expected to be decoupled from the
fractal dimension of the atomistic chain, because entanglements
arise from inter--chain rather than intra--chain configurational
properties. The estimator we develop in the following section will,
in fact, potentially be applicable to non-Gaussian chains where
$\ave{\Ree^2}\propto \epsilon^{-\mu}$ (with $1\le \mu \le 2$), as
well as less-flexible polymers (like actin \cite{actin} or
dendronized polymers \cite{dendronized}) for which $\Ne$ is
\cite{uchidaa2008} of the order of a ``persistence length" of the
atomistic chain.

\subsection{The M--kink estimator} \label{subsec:kinks}

Beyond some a priori unknown chain length $N_1$, we know that $\ZZ$
(as determined via Z1 or CReTA) varies linearly with $N$, i.\ e.\
$\ZZ=GN+Z_0$ (with $G > 0$, and $Z_0 \equiv -(G+H) > -1$ in the
notation of Eq.\ \ref{Zepsilon1}). We recall that an ideal
$\Ne$--estimator implies, according to condition (iii), that
\begin{itemize}
\item[(vi)]  $d\Nee(N)/dN=0$ for $N\ge N_1$, and
\item[(vii)] $N_1 < \Ne$
\end{itemize}
are necessary to produce $\Ne= \Nee(N_1)$. Uniquely, $\Ne = 1/G$ and $\Nee(N)=\Ne$
for all $N>N_1$. Using the linear relationship between $\ave{Z}$ and
$N$ we thus propose (a) $\Nee(N)=N/(\ZZ-Z_0)$, where $Z_0=Z_0(N)$ is
the coefficient determined from data collected up to chain length
$N$. Note that (a) is identical with the $\Ne$--estimator suggested
on mathematical grounds at the beginning of this section.

However, $\Nee(N) = 1/G$ can be equivalently obtained from (b)
$\Nee(N)=dN/d\ZZ$. This is an estimator, denoted as ``M-kink'', of
extraordinary simplicity:
\be
 \frac{1}{\Nee(N)} = \frac{d\ZZ}{dN}.
\label{Neestimatorkink} \ee M-kink is strictly an ideal estimator
(i.\ e.\ it satisfies all five conditions proposed in Section
\ref{sec:prelim}) provided $N_1 < N_e$. It eliminates the unknown
coefficient in the linear relationship, and identifies $\Ne$ to be
responsible for the ultimate \textit{slope} of $\ZZ(N)$. This is
analogous with measurements of diffusion coefficients, where one
eliminates ballistic and other contributions by taking a derivative.
Application of Eq.\ (\ref{Neestimatorkink}) requires studying more
than a single chain length, which renders our M-kink estimator
qualitatively different from the S-kink estimators. Data for
$\ZZ(N)$ for both polymer models, shown in Fig.\
\ref{Fig_N_versus_Z}, demonstrates that $\ZZ$ in fact becomes linear
in $N$ for $\ZZ$ below unity \cite{foteinopoulou08PE,footlinearZN},
thus confirming $N_1<\Ne$. This suggests that $\Ne$ can be estimated
using data for $\ZZ$ from chains of lengths even below $\Ne$.

The occurrence of a nonvanishing $N_1$ is rooted in the fact that a
minimum polymeric contour length (of the order of $2\pi \ell$ with
polymer thickness $\ell$, subsequently corrected by chemical
details) is needed for geometrical reasons to form an entanglement
(or tight knot) \cite{buck2004}. This length ($\ell$) increases with the persistence
length of the atomistic contour, and vanishes in the limit of
infinitely thin polymers. This implies that determining $\Ne$ from
the slope we correct for a thickness effect, and $N_1$ is
proportional to the thickness of the atomistic polymer.

\subsection{The M--coil estimator}
\label{subsec:coils}

Next, we motivate and derive a near--ideal estimator for use with
coil-properties $\ave{R_{ee}^2}$ and $\ave{\Lpp}$ (obtained via PPA,
CReTA or Z1). Flory's characteristic ratio $C(N)$ is defined through
the identity \cite{FloryBook,RubinsteinColby} \be \ave{\Ree^2}
\equiv (N-1)l_0^2C(N). \label{defCN} \ee Equation\ (\ref{defCN}) is
exact by construction; the $N$--dependence of $C(N)$ characterizes
the (non)-Gaussian structure of chains. In general, $C(N)\ge 1$ if
$N > 1$. For (mathematically) ideal chains, $C(N)$ is related to the
persistence length $l_p$ \cite{footCN}. This allows the chain
stiffness constant $C(\infty)\equiv \lim_{N\rightarrow \infty} C(N)$
to be calculated from short chains for any sort of ideal chain,
including random walks, freely rotating chains, wormlike chains,
etc. Simulations on dense chain packings show \cite{limitedC} that
the value of $C(\infty)=1.48$ is a universal lower limit for
excluded volume, flexible chain molecules. For real chains like
polyethylene, chains much longer than $l_p$ need to be studied to
characterize $C(N)$, cf. Ref.\ \cite{foteinopoulou08PE}. We assume
knowledge of $C(N)$ as function of $N$ from the atomistic
configurations.

\begin{figure}[tbh]
\includegraphics[width=7.5cm]{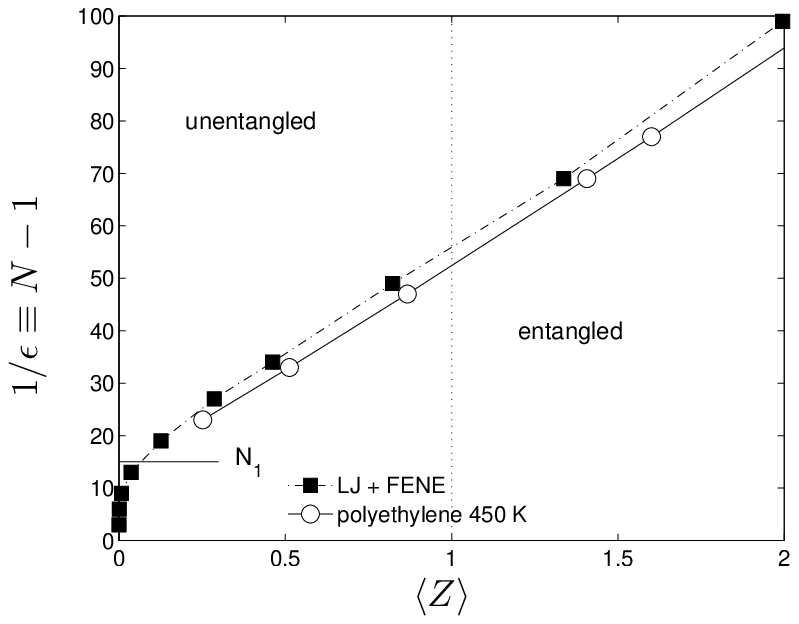}
\caption{Z1 results for the two model polymer melts. Testing the
applicability of Eq.\ (\ref{Zepsilon1}) which predicts linear
behavior in this representation (slope $G$, offset $H$). Clearly
$\ave{Z}(N)$ becomes linear at an $N$ for which $\ave{Z} < 1$. This
implies $N_1 < N_e$ and that Eq.\ (\ref{Neestimatorkink}) can be an
ideal estimator. An interpretation for $N_1$ is given in Sec.\
\ref{subsec:kinks}. Data for larger $N$ is not shown here, but the
slope $d\ZZ/dN$ does not significantly change with increasing
$N$.\label{Fig_N_versus_Z}}
\end{figure}

To proceed, we make use of our finding that $\ave{\Lpp}^2/(N-1)$ is
linear in $N$ above a certain characteristic $N_0$, before $\ZZ(N)$
has reached unity, i.e., we assume $N_0\le \Ne$ to derive an ideal
estimator (\ref{Neestimatorcoil1}). The linear relationship clearly
holds for both polymer models considered here (Ref.\
\cite{foteinopoulou08PE}, Table \ref{tab:chaindims}, Figs.\
\ref{Fig_Ree2_div_b02_Cinfty}b and \ref{Fig_N_versus_Z}), and has
already been formulated in Eq.\ (\ref{eq:LppsqNG1}). Next we relate
$\ave{\Lpp}$ and $\ave{Z}$ for large $N\gg \Ne$ by a simple
argument: the length of the primitive path, $\Lpp$, is
\cite{uchidaa2008} the number of ``entanglement nodes'', $N/\Ne$,
times the mean \textit{Euclidean} distance $\ell_e$ between such
nodes. This distance ($\ell_e$) equals the mean end--to--end
distance of the atomistic chain with $\Ne$ monomers. We thus expect
that up to a factor of order unity (related to fluctuations in $l_e$
\cite{footCN}), $\lim_{N\rightarrow \infty} \ave{\Lpp}^2 = (N/\Ne)^2
(\Ne-1) l_0^2 C(\Ne)$.

By following the procedure of Section \ref{subsec:kinks}, we arrive
at an $\Ne$--estimator, denoted as ``M--coil'', using coil
properties alone: \be
 \left(\frac{C(x)}{x}\right)_{x=\Nee(N)} =
 \frac{d}{dN} \left(
 \frac{\ave{\Lpp}^2}{R^2_\textrm{RW}}\right),
 \label{Neestimatorcoil1}
 \ee
where $R^2_\textrm{RW}\equiv (N-1) l_0^2$, and $C(x)$ is the
characteristic ratio for a chain with $\Nee(N)$ monomers. This
estimator fulfills all conditions from our above definition of an
ideal estimator. As for M--kink, the derivative in the M--coil Eq.\
(\ref{Neestimatorcoil1}) signals that we have to measure
$\ave{\Lpp}$ as function of $N$ rather than a single value to
estimate $\Ne$. The convergence properties are not as clear
\textit{a priori} as they are for the M-kink estimator Eq.\
(\ref{Neestimatorkink}), as this derivation required an
approximation. In practice, one must simulate systems with
increasing $N$ until the M--coil converges. There is no apparent way
to come up with an $\Ne$--estimator from \textit{coil} quantities
which converges before $N$ reaches $\Ne$. This is a noticeable
difference between the estimators from coils and kinks (M--kink).
Technical considerations in the application of Eq.\
(\ref{Neestimatorcoil1}) are discussed in Appendix
\ref{app:techMcoil}.

\section{Numerical results and discussion}
\label{sec:performance}

\begin{figure}[tbh]
\includegraphics[width=7.5cm]{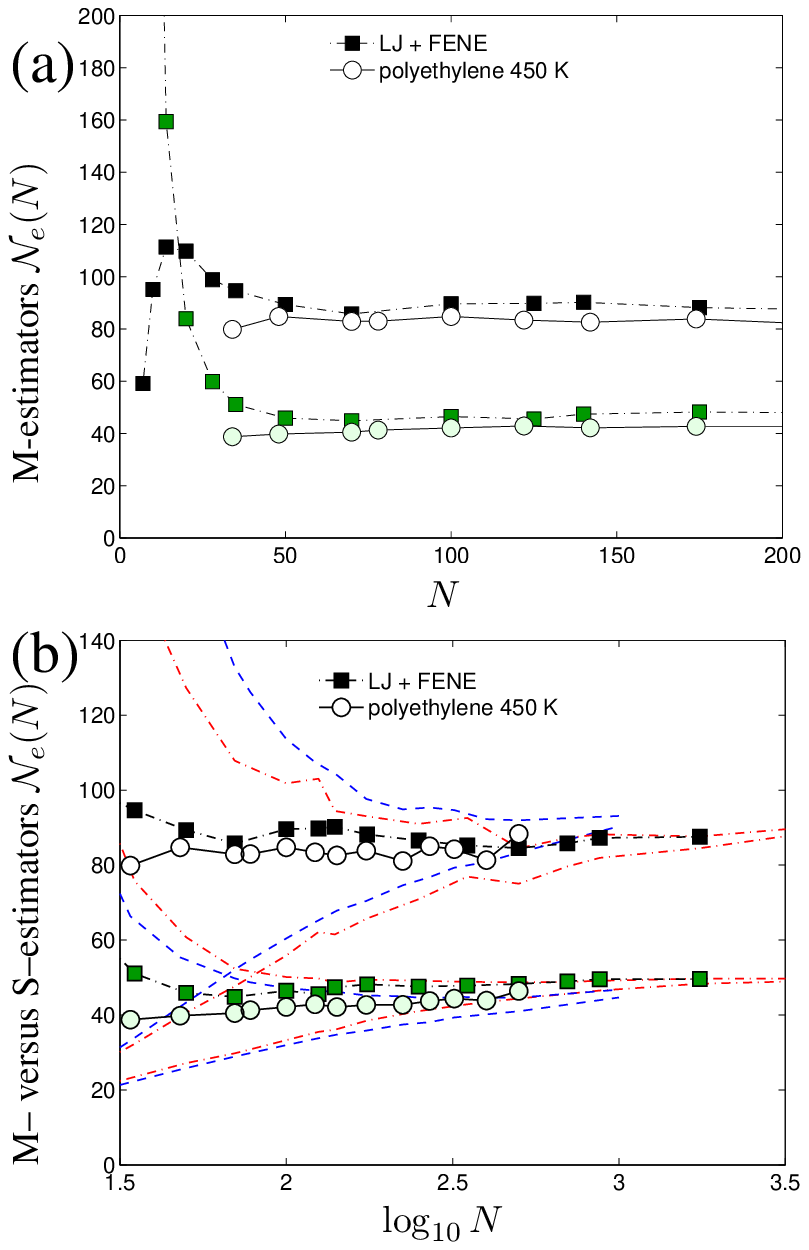}
\caption{(color online) (a) Performance of proposed estimators
M--kink (\ref{Neestimatorkink}) (lower two curves with large
symbols) and M--coil (\ref{Neestimatorcoil1}) (upper two curves with
large symbols); see also Appendix \ref{app:techMcoil}. Data is for
the same systems analyzed in Figs.\
\ref{Fig_old_standard_estimators} and
\ref{Fig_old_corrected_estimators}. Clearly, $\Nee(N)$ has converged
for $N\ll 100$, and as shown by comparison to Fig.
\ref{Fig_N_versus_Z}, $\Nee(N)$ approaches $\Ne$ before $\ave{Z}$
exceeds unity. This allows us to estimate $\Ne$ from mostly
unentangled systems. (b) Same data as in (a) vs. $\log_{10} N$,
which allows the full range of $N$ to be presented. For comparison,
blue broken and red dashed lines for PE and LJ+FENE, respectively,
show reference data for S--estimators, already presented in Figs.\
\ref{Fig_old_standard_estimators} and
\ref{Fig_old_corrected_estimators}.\label{Fig_new_standard_estimatorsDC}}
\end{figure}

The data in Tab.\ \ref{tab:chaindims} and a similar set for
atomistic polyethylene (configurations from Ref.\
\cite{foteinopoulou08PE}), will now be used to test the
M--estimators. Figure \ref{Fig_new_standard_estimatorsDC} shows
results for the M-kink estimator (Eq.\ \ref{Neestimatorkink}) and
M-coil estimator (Eq. \ref{Neestimatorcoil1}) for the same systems
analyzed in Figs.\ \ref{Fig_old_standard_estimators} and
\ref{Fig_old_corrected_estimators}. Comparison of these figures
shows that the M-- estimators indeed converge faster than the S--
estimators (Eqs.\ \ref{eq:stdEveraers}--\ref{eq:correctedEveraers}).
Moreover, comparison to Fig. \ref{Fig_N_versus_Z} shows that the M--
estimators converge for marginally entangled systems; values of
$\Nee(N)$ approach $\Ne$ before $\ave{Z}$ far exceeds unity. These
show that Eqs.\ (\ref{Neestimatorkink}) and (\ref{Neestimatorcoil1})
are essentially ``ideal'', meeting all of conditions (i)-(v). The
kink estimator performs slightly better, presumably because of the
approximations made in deriving Eq.\ (\ref{Neestimatorcoil1}).

For $N<50$, values of $\Nee(N)$  from M--kink
(\ref{Neestimatorkink}) increase with decreasing $N$. As shown in
Fig.\ \ref{Fig_new_standard_estimatorsDC}, $\Nee(N)$ appears to be
diverging as $N \to 0$. The precise nature of the divergence is
unimportant. For example, $N=20$ chains have $\ave{Z}=0.127$, and
the vast majority have zero entanglements, so the prediction
$\Nee(20) = 192 \gg 20$ of modified S--kink
(\ref{eq:correctedKroger}) just signals that we are deep in the
unentangled regime, where $\Ne$ cannot yet be estimated.

\TABLENeNe

\TABLEEARLIERNeNe

The fast convergence of the M-kink estimator can be better
understood by plugging Eq.\ (\ref{Zepsilon1}) into M--kink
(\ref{Neestimatorkink}). This produces a special case of the M--kink
estimator, which is only asymptotically correct, and can be used
when (\ref{Zepsilon1}) holds. We refer to it as the ``approximate
M--kink'' estimator: \bea \Nee(N) &\approx& \frac{1}{G}.
\label{NelinearZ} \eea Here, $G$ is the coefficient in the linear
relationship between $\ZZ$ and $N$ obtained from data collected up
to chain length $N$, and thus $\Nee(N)$ depends on $N$. Note that
the derivative with respect to $N$ in Eq. (\ref{Neestimatorkink})
removes the $O(\epsilon)$ errors! This is a major difference with
respect to all S--estimators (the estimator used in
\cite{sukumaranlong} can be considered as intermediate between S--
and M--estimators).

In a similar attempt to rationalize the fast convergence of the
M--coil estimator, we insert Eqs.\ (\ref{eq:reesqNG1}) and
(\ref{eq:LppsqNG1}) into Eq.\ (\ref{Neestimatorcoil1}).  This
yields, accordingly, the "approximate M--coil" estimator,
\bea
 \Nee(N) &\approx& 1 + \frac{D + \sqrt{ D^2-4AY}}{2A}.
\label{eq:expNeestimatorcoil2} \eea Like Eq.\ (\ref{NelinearZ}),
Eq.\ (\ref{eq:expNeestimatorcoil2}) has no $O(\epsilon)$
corrections. Again, this arises from the ``M'' approach of taking
derivatives with respect to $N$. In both cases, the use of the
derivatives removes undesirable effects related to proper treatment
of chain ends. The approximate M--coil estimator is related only to
the (in general, non-Gaussian) structure of chains and primitive
paths. Finally, if the assumptions which lead to Eq.\
(\ref{ecoils1}) hold, and in order to quantify the contributions to
Eq.\ (\ref{eq:expNeestimatorcoil2}), the above analysis combined
with tube-theoretic considerations suggests another estimator, which
we refer to as ``simplified M--coil'': \be
 \Nee(N) \approx \displaystyle\frac{D}{A}.
\label{eq:expconvstdEv} \ee The only dependence on $N$ of the
approximate and simplified estimators, Eqs.\
(\ref{NelinearZ})--(\ref{eq:expconvstdEv}), stems from the variation
of $A$, $D$, $G$, and $Y$ with $N$; these coefficients, which are
obtained by linear interpolation, must generally be assumed be
considered to depend on the available range of studied chain
lengths. When the variation of the coefficients is large, these
three estimators should not be used.

Note that the simplified M-coil does not agree with M-coil if
$C(\Ne)$ has not reached $C(\infty)$; though it may converge
quickly, it cannot be ideal. For the systems under study, $\Ne$ is
large enough such that $C(\Ne)$ is quite close to $C(\infty)$
\cite{counterexample}. The simplified M-coil has a simple connection
to polymer structure and the tube model \cite{doibook}. $D =
C(\infty) l_0^2 = l_0l_K$, where $l_K$ is the Kuhn length
\cite{footCN}. The tube diameter $d_T$ is given by $d_T^2 = l_0 l_K
\Ne$, and hence $A=(d_T/\Ne)^2$.

Table \ref{tabNeNe} quantifies the performance of the new
M--estimators. The two presented values for each estimator $\Nee(N)$
are the final $\Ne$, obtained by analyzing all available chain
lengths, together with the value predicted by the estimator at
$N=\Ne$ (i.\ e.\ at the border between unentangled and entangled
regimes, using only chains of length up to $\sim\Ne$). For an ideal
M--estimator these two numbers should be the same within statistical
errors, here $\sim 2.5\%$, and $\Ne$ should coincide with
$\lim_{N\rightarrow\infty}N/\ZZ$. All four M--estimators considered
here, the complete ones (Eqs.\ \ref{Neestimatorkink},
\ref{Neestimatorcoil1}) as well as their approximate versions (Eq.\
\ref{NelinearZ}, \ref{eq:expNeestimatorcoil2})  satisfy these
criteria. The simplified M--coil (\ref{eq:expconvstdEv}) is seen to
converge quickly as well, but it does converge to an $\Ne$ which is
above the one obtained via M--coil, because $Y$ is positive ($Y$
vanishes for an ideal random walk). Table \ref{tabNeNeearlier} shows
corresponding results for the S--estimators, which all (as discussed
above) are generally non-ideal. Still, the modified S--kink turns
out to perform very well, simply because $G+H\ll 1$ for our model
systems, cf.\ Tab.\ \ref{tabDYAC2}.

For the LJ+FENE model, while the classical S--coil estimator (Eq.\
\ref{eq:stdEveraers}) produces values of $\Ne$ consistent with
published \cite{sukumaranlong} results, i.\ e.\ $\Ne \simeq 70$ for
$N = 350$ and $500$, values for these estimates based on the
near--ideal M--estimators (cf. Tab. \ref{tabNeNe}) and also the
modified S--coil (\ref{eq:correctedEveraers}) rise above $80$ for
the longest chains considered here. The M--estimators based on chain
and (Z1) primitive path dimensions converge to the value $\Ne \simeq
85$ in the mostly unentangled regime, cf. Tab.\ \ref{tabNeNe}. Thus
all data suggest that the ``best'' estimate of the entanglement
length for flexible chains is well above the previously reported
value. This is significant e.\ g,\ for quantifying the ratio
$N_e/N_c$, where $N_c$ is the rheological crossover chain length
where zero shear viscosity changes its scaling behavior from Rouse
to reptation, and has been estimated as $N_c \approx 100$
\cite{kh1995,kh2000}.

One could imagine fitting the squared contour length
$\ave{\Lpp^2(n)}$ of  primitive path subsections \cite{footL} to
$\ave{\Lpp^2(n)} = An^2 + Cn$.
and attempting to calculate $\Nee(N) = D/A$ by also fitting to
$\ave{R^{2}(n)} = Dn - Y$, or developing other improved estimators
for $\Ne$ based on $\ave{\Ree^2(n)}$ and $\ave{\Lpp^2(n)}$. However,
analysis along these lines failed to produce any estimators better
than those described above. In particular, no improvement over the
method of Ref.\ \cite{sukumaranlong} was found.

It is important to notice that our Eq.\ (\ref{Neestimatorcoil1}) is
\textit{not} compatible with some earlier definitions of $\ZZ$ from
coil quantities, because of the prefactor $C(\infty)/C(\Ne)$. This
prefactor had usually been omitted or not mentioned, since random
walk statistics were clearly a convincing starting point. Assuming
Gaussian statistics (constant $C(N)$ for all $N$) hence
underestimates values of $\Ne$ calculated from coil properties. This
issue is also one of the reasons why the $\Ne$ estimates between PPA
and geometrical approaches differ. Another reason is given in
\cite{elastic}. Ratios between 1.3 and 2.5 between $\Ne$ calculated
from kinks and coils have been reported in the literature
\cite{christos,Foteinopoulou2006,christoscurropin,cpc}. The
presented data exhibits ratios between 1.6 and 2. A third reason
that they differ is rooted in the fact that $\ZZ$ is not \cite{cpc}
uniquely defined from a given shortest, piecewise straight path, as
it is returned by Z1 or CReTA. This additional discrepancy can only
be resolved by matching results for $\Ne$ from kinks and coils, and
by comparison with experiments.

The classical S--kink (\ref{eq:stdKroger}) strictly underestimates
$\Ne$ and the modified S--kink (\ref{eq:correctedKroger}) strictly
overestimates $\Ne$ (since both $G$ and $H$ are positive, and
$G+H<1$).

\section{Conclusions}
 \label{sec:conclude}

Very significantly improved, near--ideal, and apparently
polymer--model--independent estimators for $\Ne$ were derived in
this paper, M--coil (Eq.\ \ref{Neestimatorcoil1}, to be used with
PPA, Z1 or CReTA) and M--kink (Eq.\ \ref{Neestimatorkink}, Z1 and
CReTA only). They reduce, under further assumptions which seem valid
for the model systems studied here, to approximate M--coil (Eq.\
\ref{eq:expNeestimatorcoil2}), simplified M--coil (Eq,\
\ref{eq:expconvstdEv}), and approximate M--kink (Eq.\
\ref{NelinearZ}). These estimators require simulation of multiple
chain lengths, but have eliminated systematic $O(\epsilon)$ errors
present in previous methods. This is important for the design of
efficient simulation methods in the field of multiscale modeling of
polymer melts.

Furthermore, we have proposed variants of the original estimators.
The two main problems with existing estimators were identified as:
i) improper treatment of chain ends, and ii) nontreatment of the
non-Gaussian statistics of chains and primitive paths
\cite{alexpriv}. Improper handling of thermal fluctuations was an
additional problem relevant to very short chains. Issues i) and ii)
lead to separate, independent $O(\epsilon)$ errors. Estimators based
on direct enumeration of entanglements lack issue ii), and so are
\textit{fundamentally} advantageous for estimation of $\Ne$. The new
``M'' estimators proposed here formally correct for the errors
arising from effects i) and ii). The values of the M--coil and
M--kink--estimators can be taken as ``best estimates'' for $\Ne$
when results are available for multiple chain lengths. The best
estimator when only a single chain length is available is the
modified S--kink, Eq.\ (\ref{eq:correctedKroger}).

We have shown that $\epsilon\ave{\Lpp}^2$, $\ave{Z}$, and also
$\ave{\Ree^2}$ are all linear in $1/\epsilon$ (thus linear in $N$)
down to the mostly unentangled regime, and have used
this information to derive the M--estimators and to improve the
earlier ones. All coefficients in these linear relationships have
been evaluated and listed in Tab.\ \ref{tabDYAC2}. The prefactors
for the above mentioned $O(\epsilon)$ errors can be large, and
depend both on the polymer model and method of topological analysis.
These errors can produce large changes in estimates of $\Ne$ for
values of $N$ typically considered in previous studies (e.\ g.\
Refs.\ \cite{everscience,shanbhag06,leon05}). This is significant in
light of attempts to compare PPA results for $\Ne$ to values
obtained by other methods
\cite{putz00,everscience,leon05,likhtman07} such as direct
rheological measurement of the plateau modulus $G_N^0$, evolution of
the time-dependent structure factor $S(\vec{q},t)$, and estimation
of the disentanglement time $\tau_{d} \propto (N/\Ne)^3$
\cite{doibook}. Some conclusions of those studies may need to be
reevaluated in light of the new data.

The proposed M--estimators are to our knowledge the first estimators
which exhibit all features required for an ideal estimator (a term
which we made precise in Sec.\ \ref{sec:methodsdims}), and they have
been physically motivated. They converge to $\Ne$ for weakly
entangled systems ($N\le \Ne$). They leave $\Ne$ either undefined or
infinite for rodlike chains (because $C(N)=N$ for a rod). They
predict $\Nee(N)\ge N$ for a completely unentangled system, which is
characterized by $\ZZ=0$ and $\Lpp=\Ree$ in accord with the
definition of the primitive path which we have adopted in this work
(see \cite{elastic}). The appearance of the coefficient $N_1$
suggests that there might be a minimum amount of material, $N_1$,
needed to form a single entanglement (as observed for phantom chains
\cite{cpc}). If so, it can be expected to depend on the thickness of
the atomistic chain and its stiffness as well as particle density.
We expect our findings to be universal in the sense that they
should apply to all sorts of real linear polymer chains in the melt
state, and we have verified the assumptions underlying the
M--estimators by direct comparison with both atomistic semiflexible
and coarse-grained flexible polymer melts.

Refs.\ \cite{christos,christoscurropin} pointed out that primitive
paths are not random walks, and  that there appears to be more than
one ``topological'' entanglement per ``rheological'' entanglement;
thus it is unsurprising that $\Ne$ from coils is significantly
larger than $\Ne$ from kinks (for details see Ref.\
\cite{uchidaa2008}). The utility of any topological analysis of
chains shorter than $\Ne$ remains highly questionable, because the
chains' dynamics are well described by the Rouse model
\cite{doibook,kremer90} and so they cannot be considered
``fully entangled'' in any meaningful way. However, it seems that the
M--estimators developed in this work have the ability to extract information from a
partial or even marginal degree of entanglement.

The $M$--estimators could be applied in a post-processing step on
existing configurations. For example, it should be of interest to
study the effect of flow and deformation on entanglement network
characteristics in order to establish equations of motion for
relevant coarse-grained variables characterizing the polymer melt.
Shear and elongational flows have been studied for both polymer
models considered here, but either Z1 was not yet available at the
time of these studies \cite{loose}, or the chains were
\cite{keffer,cochran} ``too short'', i.\ e.\ had $\ZZ\ll 1$.

The apparent ability to accurately estimate $\Ne$ even for weakly
entangled systems may be useful for atomistic models whose
computational cost prohibits equilibrating large-$N$ systems, such
as polymers containing bulky side groups. The procedure for removal
of the $O(\epsilon)$ systematic errors, while clearly described
here, requires performing analyses on a limited number of
configurations on a range of chain lengths, which is most easily
undertaken for systems composed of ``short, but not too short''
chains.

\section{Acknowledgements}

R.\ H.\ thanks Alexei E.\ Likhtman for pointing out that the
non-Gaussian statistics of chains and primitive paths produce
$O(1/N)$ systematic errors in the old estimators for $\Ne$. Steven
J.\ Plimpton integrated the DBH algorithm into LAMMPS \cite{LAMMPS}.
Gary S.\ Grest and Nikos Karayiannis provided
helpful discussions.  Gary also provided an equilibrated
$N=3500$ state, and Nikos was deeply involved in all PE developments.
This work was supported by the MRSEC Program of the
National Science Foundation under Award No.\ DMR05-20415, as well as
through EU-NSF contract NMP3-CT-2005-016375 and FP6-2004-NMP-TI-4
STRP 033339 of the European Community. All atomistic simulations
were conducted in the ``magerit'' supercomputer of CeSViMa (UPM, Spain).

\begin{appendix}

\section{Treatment of Thermal Fluctuations} \label{app:fluctuations}

Ref.\ \cite{everscience} and other studies have typically used
$\ave{\Lpp}^2$ rather than $\ave{\Lpp^2}$ in estimators for $\Ne$,
such as the analogue for the modified S-coil
(\ref{eq:correctedEveraers}) which reads
\begin{equation}
 \Nee(N) = (N-1)\left(\displaystyle\frac{\ave{\Lpp}^2}{\ave{\Ree^2}} - 1\right)^{-1}.
 \label{eq:mangledEveraers}
\end{equation}
However, Eq.\ (\ref{eq:mangledEveraers}) gives pathological results
for short chains due to existing thermal fluctuations of $\Lpp$.
Consider the unentangled limit, where the entanglement density
(denoted as $\rho_e$) vanishes. For an ``ideal'' topological
analysis, $\Lpp \to \Ree$ (from above) for each and every chain as
$\rho_e\rightarrow 0$. However, chain dimensions fluctuate in
thermodynamic equilibrium \cite{doibook}. To leading order in the
fluctuations, $\ave{\Lpp}^2 = \ave{\Lpp^2} - (\Delta \Lpp)^2 \equiv
\ave{\Ree^2} - (\Delta \Ree)^2$, where $\Delta$ is ``variance of''.
So, even for an \textit{ideal} topological analysis procedure, Eq.\
(\ref{eq:mangledEveraers}) would predict a \textit{negative}
$\Nee(N) \to -(N-1)\ave{\Ree^2}/(\Delta \Ree)^2$ as $\rho_e \to 0$.
Negative $\Nee(N)$ are of course useless, but indeed, are predicted
using our data in Tab.\ \ref{tab:chaindims}. For $N=20$ (LJ+FENE
melt), application of Eq.\ (\ref{eq:mangledEveraers}) yields
negative $\Nee(20)$. A term identical to the term in parenthesis in Eq,\ \ref{eq:mangledEveraers} was  found to be negative for short chains in Ref.\ \cite{keffer}, but was not used to directly calculate $\Nee(N)$ in their work, as its negative value was considered to signal (and to only occur in) the mostly unentangled regime.

The reason to fix chain ends during PPA or Z1 analysis is the
assumption, implicit in Edwards' definition of the primitive path
\cite{edwards77}, that chains are entangled. In this context it is
worthwhile mentioning that there are other definitions of PP's, for
example one \cite{ramirez} where the length of the PP goes down to
zero for the unentangled chain, and where chain ends are not fixed.

\section{Technical Considerations in Use of the M--Coil Estimator}
\label{app:techMcoil}

\begin{figure}[tbhp]
\includegraphics[width=7.5cm]{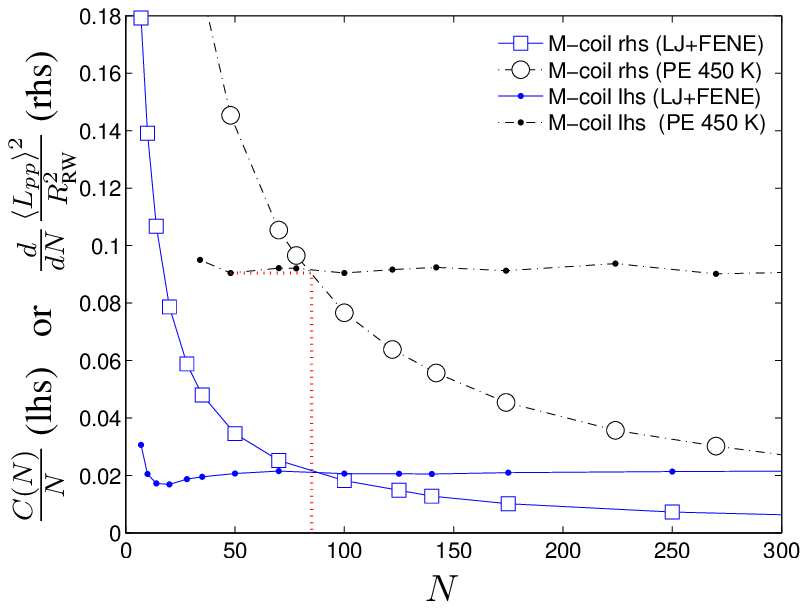}
\caption{The graph demonstrates on how to graphically evaluate
$\Nee(N)$ according to the M--coil estimator
(\ref{Neestimatorcoil1}). Shown are both the left hand side (lhs)
$C(N)/N$, and right hand side (rhs) of Eq.\ (\ref{Neestimatorcoil1})
for both types of polymer melts. The dotted red path makes an
example on how to obtain $\Nee(N)\approx 87$ for given $N=48$.
Obviously, this value is quite identical with both, $\Ne$ and
$\Nee(\Ne)$, cf. Tab.\ \ref{tabNeNe}.
The ratio $C(N)/N$ (small points) monotonically decreases with
increasing $N$, while the rhs (large symbols) reaches a plateau at
the time $N$ has approached $\Ne$ (at the crosspoint), which is a
distinguished feature of an ideal
estimator.\label{Fig_Neestimatorcoil_both_sides}}
\end{figure}

While the M--kink estimator (Eq.\ \ref{Neestimatorkink}) is
explicitly evaluated from the local derivative $d\ave{Z}/dN$ around
$N$, our M--coil expression, Eq.\ (\ref{Neestimatorcoil1}), is only
an implicit expression for the estimator $\Nee(N)$. Formally, we
need the inverse of $C(N)/N$ to calculate $\Nee(N)$. In the
following, we describe the procedure in order to prevent any
ambiguities upon applying M--coil in practice. Fig.\ \ref{Fig_Neestimatorcoil_both_sides} shows both the  left (lhs) and right hand (rhs) sides of Eq.\ \ref{Neestimatorkink}
versus $N$ for our data. 
For any given $N$ (say, $N=48$ for the PE data, where the dotted red line
starts in Fig.\ \ref{Fig_Neestimatorcoil_both_sides}), the $\Nee(N)$
estimate is the value at the ordinate for which the abscissa values
for lhs and rhs coincide (end of the red curve is at
$\Nee(48)\approx 87$). The same procedure is repeated for all $N$ to
arrive at Fig.\ \ref{Fig_new_standard_estimatorsDC} and particular
values collected in the M--coil row of Tab.\ \ref{tabNeNe}.
The difference between lhs and rhs can be used to
estimate the difference between the largest $N$ available and $\Ne$.
If only short chains had been studied, only a part of
this plot could have been drawn. 

Note that this procedure requires $C(N)/N$ to be monotonically
decreasing with $N$, and access to $C(N)$ at sufficiently large $N$.
While the former is essentially valid for all polymer models, the latter may pose a problem.
Without reliable values for $C(N)$ for $N=\Ne$, there is no apparent way to come
up with an M--coil which converges before $N$ reaches $\Ne$. 
However, since $C(N)/N$ decreases with increasing $N$ and ultimately reaches $C(\infty)/N$ behavior, in practice (and formally for ideal chains) $C(N)$ can be estimated by extrapolation, and the necessary $C(N)/N$ values could be added for chain lengths exceeding those studied.

This issue disappears by construction when the largest simulated $N$ exceed $\Nee(N)$, so that the conditions for an ideal estimator are met in any case. 
Still, this is a noticeable and principal difference
between the estimators from coils (M--coil) and kinks (M--kink).

\end{appendix}


\begin{thebibliography}{99}
\bibitem{footCN}
For the case of wormlike (non-gaussian) chains, $C(N)$ is
analytically related to the persistence length $l_p$ of short
chains; $C(N+1) = C(\infty) - 2\alpha(1-\alpha^N)(1-\alpha)^{-2}/N$,
where $\alpha=\exp(-l_0/l_p)$, and $C(\infty)=\lim_{N\rightarrow
\infty}C(N) = (\alpha+1)/(\alpha-1)$. Note that
$\lim_{\alpha\rightarrow1}C(N)=N$ and $\lim_{\alpha\rightarrow
0}C(N) =1$. In the limit $l_p\gg l_0$, the mean square end--to--end
distance of a wormlike chain consequently reads $\ave{\Ree^2}=2l_p
L(1-[l_p/L][1-\exp(-L/l_p)])$ with contour length $L\equiv
(N-1)l_0$. In the further limit $L\gg l_p$, $\ave{\Ree^2}/(N-1)=2l_p
l_0 = C(\infty) l_0^2\equiv l_K l_0$, where $l_K$ denotes Kuhn
length. These limits are not employed in this work, as we are
seeking for corrections in the regime where these limits have not
necessarily been reached. For freely rotating chains with a fixed
bending angle $\theta$ the same result is obtained upon identifying
$\alpha=\ave{\cos\theta}$. The above expressions for $C(N)$ are
directly obtained from a bond--vector correlation function which
exponentially decreases with distance between bonds. However,
simulations have shown \cite{everaers} that flexible bead spring
chains in melts do not follow this form at small and moderate $N$. The ratio
$C(N)$ must sublinearly and monotonically increase with $N$ in order
to uniquely determine $\Ne$ from (\ref{Neestimatorcoil1}), which
limits its use to systems which do not change their statistics on a
length scale larger than entanglement length. For the wormlike chain
as well as equilibrated bead spring chains \cite{everaers}, $C(N)/N$
decreases monotonically with $N$.
\bibitem{footlinearZN}
Ref.~\cite{foteinopoulou08PE} reported $\Ne>40$ and that $\ave{Z}$
to became linear at $N>24$.
\bibitem{counterexample}
A counterexample would be LJ+FENE bead spring chains made
semiflexible by the addition of a bending potential $U = -k_\theta
\cos(\theta)$; when $k_\theta=2\,\epsilon_\textrm{LJ}$, $C(\Ne)$ is
still well below $C(\infty)$ \cite{everaers}.
\end{thebibliography}
\end{document}